\documentclass[12pt,aps,prb,preprint]{revtex4}

\usepackage{amsmath}
\usepackage{graphicx}
\newcommand{\m}[1]{\lvert#1\rvert} 
\newcommand{\M}[1]{\lVert#1\rVert} 
\makeatletter
\newlength{\PM@dpth}\newlength{\PM@hght}\newlength{\PM@wdth}
\newcommand{\clx}[1]{\settodepth{\PM@dpth}{$\underline{a}$}\settoheight{\PM@hght}{$#1$}%
\addtolength{\PM@hght}{\PM@dpth}\settowidth{\PM@wdth}{$#1$}\makebox[0pt][l]{\rule[\PM@hght]%
{\PM@wdth}{\fboxrule}}#1} \makeatother
\providecommand{\pair}[2]{\ensuremath{\leval\, #1\,,\, #2\,\reval}}  
\newcommand{\leval}{{\mbox{\textnormal[\hspace{-0.17em}\textnormal[}}}
\newcommand{\reval}{{\mbox{\textnormal]\hspace{-0.17em}\textnormal]}}}

\begin{document}
\title{Effects of walls}
\author{T. B. Smith}
\affiliation{The Open University, Department of Physics and Astronomy, Walton Hall,
Milton Keynes, MK7 6AA, UK}
\email{t.b.smith@open.ac.uk}
\author{D. A. Dubin}
\affiliation{The Open University, Department of Pure Mathematics (retired),
Walton Hall, Milton Keynes, MK7 6AA, Uk}
\author{M. A. Hennings}
\affiliation{Rugby School, Rugby, Warwickshire, CV22 5EH, UK}
\date{\today}
\begin{abstract}
We analyze here the energy states and associated wave functions
available to a particle acted upon by a delta function potential
of arbitrary strength and sign and fixed anywhere within a
one-dimensional infinite well. We consider how the allowed
energies vary with the well's width and with the location of the
delta function within it. The model subtly distinguishes between
whether the delta function is located at rational or irrational
fractions of the well's width: in the former case all possible
energy eigenvalues are solutions to a straightforward dispersion
relation, but in the later case, to make up a complete set these
`ordinary' solutions must be augmented by the addition of
`nodal' states which vanish at the delta function and so do not
`see' it. Thus, although the model is a simple one, due to its
singular nature it needs a little careful analysis.  The model, of
course, can be thought of as a limit of more physical smooth
potentials which, though readily succumbing to straightforward
numerical computation, would give little generic information.
\end{abstract}
\maketitle
\section{Introduction}\label{intro}
Calculating the energy states for the motion of a particle
in a `simple' one dimensional infinite square well finds its way into all
standard quantum mechanics texts.  But because the potential
representing the walls is sharp and infinite the model has
subtleties. For instance a consistent definition of a momentum
operator for it is challenging \cite{dubin1, garcia}.
Another intriguing property of the infinite well is that for
certain initial states the time development of the spatial
probability density can show fractal behavior
\cite{berry1}. There is also the topic of wave packet
revival in such a well \cite{rob, styer}. And the time
dependence of a particle's wave function in a suddenly expanded
well can show interesting features \cite{aslangul}. Clearly, when sharp
edges and infinities are involved, care is important.

In this paper we compound the singularity by adding
within the well a fixed delta function potential of arbitrary sign, strength
and location. For this model, due to the containing walls, all solutions
to the Schr\"{o}dinger equation are bound states,
but the model does have its niceties. We find for instance that although,
as expected for the bound states of any one-dimensional system, all of its
energy eigen-states are non-degenerate, should the delta function be located at a
rational fraction of the well's width, then a subset of these energies
approach {\em in the limit} of strong attraction or repulsion, double
degeneracy.

In addition to examining several other properties and
limits of the model we have computed---for both attraction and repulsion---
the ground state energy of the system as a function of the location of the delta function.  For the case of
attraction this energy has a minimum when the force center is located at the center of the
well. For a wide box this minimum is nearly flat, but it becomes
sharper when the well width is reduced to sizes of order of the
spatial decay length characterizing the `molecular' size of the bound state of
an attractive delta function in free space.

The model has not, of course, escaped attention. Patil \cite{patil} and Atkinson and Crater
\cite{atkinson} give analyses for the case that the delta function potential is
located precisely at the center of the well. Bera and his
co-workers \cite{bera} place it anywhere within the
well and utilize perturbation theory. Joglekar \cite{jog} computes
the energy eigenvalues when it is located at
irrational multiples of the well's width and considers weak and
strong coupling limits. Epstein \cite{epstein} uses the well with
the delta function located at the center in order to gain insight into
the interesting question posed by Wigner \cite{wigner1} as to
whether the energy levels of a hydrogen atom should be derivable
from second order perturbation theory in view of the fact that it
is known to be proportional to the fourth power of the electron
charge and therefore to the square of the Coulomb interaction.

We were motivated to consider the model by the long standing
problem of the effects of containment on molecule states and energies
dating back to work by Michels \cite{michels}, Sommerfeld \cite{sommerfeld} and their co-workers, in
the early heyday of quantum mechanics, applying approximations to study what might be said
to be the canonical problem---that of a hydrogen atom with its nucleus
fixed at the center of an impenetrable sphere. Other more recent
papers discuss this model with various methods of approximation.  See for instance references \cite{varshni1,
varshni2, yngve, goldman, goodfriend} and, especially \cite{froeman}
which includes a short review of work on this model up to 1987.  More recently Changa and
co-workers \cite{changa} have applied perturbation theory to
estimate the energies when the hydrogen nucleus is {\em shifted} from
the center of the the spherical box.

The particular advantage of the model we consider here is that no approximations are
necessary to discuss the energy levels and their dependence either on the
well width or on the location of the force center. It would be gratifying to be able to do
the same with other model interactions. For instance one might try
a `one dimensional Coulomb' repulsion or attraction, $v(x) \sim 1/|x|$, $x\neq 0$.
But it is a sobering fact that even with no walls present, so that $x$ is any real number
bar zero, there seems to be no final agreement for that model as to the available energy levels
and corresponding eigenstates \cite{loudon, andrews, xdd, nouri, a a}.

The paper is arranged as follows.
We give the general solution in Sec.~\ref{solution}. Its nature depends upon
whether the ratio $\ell/L$ is a rational or irrational number, where $\ell$ is
the delta function's position within a well of  width $L$.
When that ratio is irrational all solutions are straightforward `ordinary'
solutions, but when it is rational a complete set of solutions
includes both the ordinary ones together with those we choose to call
`nodal' solutions. For both cases we confirm in Sec.~\ref{orthog} mutual
orthogonality between all solutions corresponding to different
energies. In Secs. \ref{weak} and \ref{strong} we consider the limits of weak and strong
coupling for any location of the delta function
potential within the well. In taking these limits, when $\ell/L$
is rational care must be exercised to include both the nodal and
ordinary solutions. In Sec.~\ref{1/2} we discuss solutions for the
special symmetric case that the delta function is located
precisely in the middle of the well, for which $\ell/L$ is the
rational number $1/2$. In Sec.~\ref{gs} we consider in some detail the ground state of the
model, especially with respect to its energy dependence on the
relative location $\ell/L$, for various signs and strengths of the
delta function potential. This is plotted in Fig.~\ref{single}.
\section{solution}\label{solution}
A particle of mass $m$ free to move within a one-dimensional
well with sides at $x = 0$ and $x=L$ has energy eigenstates
\begin{equation}\label{well-1}
  \psi^{(0)} _N (x) = \sqrt{\frac{2}{L}}\,\sin\left (\frac{N \pi
  x}{L}\right), \quad N=1,2,3,\ldots\,,
\end{equation}
and corresponding energies
\begin{equation}\label{energy-1}
E^{(0)}_N = \frac{\hbar^2}{2m} \left(\frac{N \pi}{L}\right)^2.
\end{equation}
For odd values of $N$ these states are symmetric about the middle
($x=L/2$) while those for even $N$ are antisymmetric.

A particle constrained by no walls but acted upon by a
one-dimensional delta function potential located at $x=\ell$, namely
\begin{equation}\label{deltaf}
v(x) = - \lambda \delta(x-\ell),
\end{equation}
has the associated Schr\"{o}dinger equation
\begin{equation}\label{schrod}
-\frac{\hbar^2}{2m}\Psi''(x) - \lambda \delta(x- \ell)\Psi(x) = E \Psi(x).
\end{equation}
where, if $\Psi_-$ and $\Psi_+$ are solutions to the left
($x<\ell$) and right ($x>\ell$) of the delta function, the
conditions to be satisfied at the potential are
\begin{equation}\label{bc-1}
-\frac{\hbar^2}{2m}\left( \Psi'_+(\ell) - \Psi'_-(\ell) \right)=
\lambda \Psi_-(\ell)\quad{\rm and}\quad \Psi_-(\ell)=\Psi_+(\ell).
\end{equation}
For an attractive interaction ($\lambda>0$) the single bound state is
represented by
\begin{equation}\label{bc-11}
\Psi_\ell\,(x)= \frac{1}{\sqrt{\Lambda}}\, {\rm exp}\left({-\frac{|x-\ell|}{\Lambda}}\right),
\end{equation}
where
\begin{equation}\label{biglam}
    \Lambda \equiv \frac{\hbar^2}{m \lambda}
\end{equation}
is a measure of the `size' of the bound state and the energy is
\begin{equation}\label{ebee}
-E_B \equiv -\frac{m \lambda^2}{2 \hbar^2} = -\frac{\hbar^2}{2 m} \frac{1}{\Lambda^2}.
\end{equation}
Whichever the sign of $\lambda$, solutions to (\ref{schrod}) also include a continuum of positive
energy unbound states, the scattering states.

Now consider putting the potential (\ref{deltaf}) inside a box of
finite side $L$. Then all states of the system are bound under the
combined action of the walls and delta function.  For this model,
solutions must satisfy Eq. (\ref{schrod}), must vanish at the walls and
must obey conditions (\ref{bc-1}).  For $x \neq \ell$
solutions of Eqs. (\ref{deltaf}) and (\ref{schrod}) are
linear combinations of ${\rm exp}(\pm{\rm i}kx)$, with energy $E =
\hbar^2 k^2/2m$.  To be continuous at $x =
\ell$ and vanish at $x=0$ and $x= L$ solutions will have the form
\begin{equation}\label{sol-1}
    \psi(x) = \begin{cases}
   \psi_{+}(x)= C \sin(k(L-x)) \sin(k \ell) & \text{for $\ell \leq  x \leq L $}\\
   \psi_{-}(x) =   C \sin(kx) \sin(k(L- \ell)) & \text{for $0 \leq x \leq
    \ell$},
    \end{cases}
\end{equation}
where $C$ is a constant which may depend on $k$, $L$ and $\ell$.

Applying the first of boundary conditions (\ref{bc-1}) gives the
energy dispersion relation.  It is
\begin{equation}\label{dispeq}
    k \Lambda \sin(kL) = 2 \sin(k \ell) \sin(k(L-\ell)),
\end{equation}
where $\Lambda$ is given by (\ref{biglam}).
One set of solutions to (\ref{dispeq}) is given by
\begin{equation}\label{cursol}
    \sin(kL)=0,\quad\quad \sin(k \ell) =0, \quad\quad \sin(k(L-\ell))=0,
\end{equation}
where any one of these equations implies the other two. Here $k$
must meet the two-fold condition
\begin{equation}\label{sol-2}
    k = \frac{\pi n}{L} = \frac{\pi p}{\ell}\quad\mbox{so
    that}\quad \frac{\ell}{L} = \frac{p}{n},
\end{equation}
where $n$ and $p$ are positive integers.  We shall assume that the ratio
$p/n$ has been reduced to its primitive form---that is to say with
no common factors other than unity---so that, because $\ell/L =
p/n = pj/(nj)$, we can write all of these {\it nodal}
solutions $\psi^{(\textbf {n})}$, and their energies, for any such given value of
$\ell/L$ as
\begin{equation}\label{sol-3}
    \psi_{n,\,j}^{(\textbf{n})}(x) = \sqrt{\frac{2}{L}}\sin\left(\frac{n j\, \pi
    x}{L}\right)\,\,\mbox{and}\,\,
    \,\, E_{n,\,j}=
    \frac{\hbar^2}{2m}\left(\frac{nj\, \pi }{L}\right)^2, \quad
    j=1,2,3\ldots,
\end{equation}
with wave numbers given by
\begin{equation}\label{sol-31}
    kL = j n \pi.
\end{equation}
The particular feature of these solutions is that they have {\em
nodes} at the location of the delta function potential. They are
that subset of the standing wave solutions (\ref{well-1}) for a
free particle in a well with this property. Because they vanish at
$x=\ell$ they satisfy both conditions (\ref{bc-1}). As an example,
when $\ell/L$ equals $1/3$ (or $2/3$) the states are
$\psi_{3,\,j}^{(\textbf{n})}(x) = \sqrt{2/L}\,
\sin(3 \pi j x/L)$ with energies $E_{3,j} = (\hbar^2/2m)(3 \pi
j/L)^2$ where $j=1,2,3\ldots$.

When $\sin(kL)$ does not vanish Eq. (\ref{dispeq}) can be re-expressed as
\begin{equation}\label{disp-21}
    k \Lambda = k L \left(\frac{\Lambda}{L}\right) =\frac{2 \sin(k \ell)
    \sin(k(L-\ell))}{\sin(kL)},
\end{equation}
or, equivalently, as
\begin{equation}\label{LB-1}
    \frac{kL}{2} \left(\frac{2 \Lambda}{L}\right) = \tan\left(\frac{kL}{2}\right)
    -2\frac{\sin^2\left(\frac{kL}{2}\mu \right)}{\sin(kL)}\quad \mbox{where}
    \quad \mu = 2\frac{\ell}{L}-1\,
\end{equation}
and $\ell$ varies from $0$ to $L$ as $\mu$ ranges from $-1$ to $1$.  We choose to call solutions
for $k$ to either of these versions of the dispersion relation (together with their associated
wave functions) {\em ordinary} solutions.

Should $\ell/L$ be a rational number, complete information about the nodal states
and their energies (all positive) are given by Eqs. (\ref{sol-3}) and (\ref{sol-31}).
The ordinary states exist for all values of $\ell$
between $0$ and $L$, with energies $E= \hbar^2 k^2/(2 m)$.  For positive energies
the allowed values of $k$ are real-valued.  In that case, solutions for $kL$ to Eq. (\ref{disp-21})
occur at intersections with the expression on the right-hand side of a
straight line passing through the origin with slope $(\Lambda/L)$. Similar comments apply
to Eq. (\ref{LB-1}) in terms of $kL/2$ and slope $2 \Lambda/L$.  For negative energies, should there
be any, the wave number is $k= {\rm i} \kappa$, where $\kappa$ is real.  In that case the energy is $E=- \hbar^2
\kappa^2/(2 m)$ and, letting $k \rightarrow {\rm i}\kappa$ in the pair (\ref{disp-21})
and (\ref{LB-1}) and in (\ref{sol-1}), one must solve either of
the equations
\begin{equation}\label{disp-3}
\kappa L \left(\frac{ \Lambda}{L}\right) = \frac{2 \sinh(\kappa \ell)
    \sinh(\kappa(L-\ell))}{\sinh(\kappa L)}= \tanh\left(\frac{\kappa L}{2}\right)
     - 2 \frac{\sinh^2 \left(\frac{\kappa L}{2}\mu \right)}{\sinh(\kappa
     L)},
\end{equation}
with corresponding wave functions
\begin{equation}\label{sol-11}
    \psi(x) = \begin{cases}
    D \sinh(\kappa (L-x)) \sinh(\kappa \ell) & \text{for $\ell \leq  x \leq L $}\\
    D \sinh(\kappa x) \sinh(\kappa (L- \ell)) & \text{for $0 \leq x \leq
    \ell$}\,.
    \end{cases}
\end{equation}
\section{orthogonality}\label{orthog}
We choose to confirm by direct calculation that for this
one-dimensional problem all solutions corresponding to different
energies are, as expected, mutually orthogonal.

Note first that, because the nodal solutions are free particle states in a simple well, it
follows at once that those for different energies are mutually
orthogonal.  Then too these nodal solutions, Eq. (\ref{sol-3}), are orthogonal to
the ordinary states. To see this, write the nodal states in condensed notation as
\begin{equation}\label{curwf}
    \psi^{({\textbf{n}})}(x) = \sqrt{\frac{2}{L}}\sin(\nu x),\quad\mbox{where}\quad \nu =
\frac{n j\,\pi}{L}= \frac{pj\,\pi}{\ell}.
\end{equation}
Now, for $k\neq 0$ and $k \neq \nu$,
\[\int_0^\ell\sin(\nu x) \sin(kx)\,{\rm d}x = (-)^{pj} \,\frac{\nu \sin(k\ell)}{k^2-\nu^2} \]
and
\[\int_\ell^L \sin(\nu x) \sin(k(L-x))\,{\rm d}x =
-(-)^{pj}\,\frac{\nu \sin(k(L-\ell))}{k^2-\nu^2}. \]
Then using definition (\ref{sol-1}) gives
\begin{equation}\label{curorthog}
    \int_0^L \psi(x) \psi^{({\textbf{n}})}(x)\,{\rm d}x= \int_0^\ell\psi_-(x)\psi^{({\textbf{n}})}(x)\,{\rm}dx
+ \int_\ell^L \psi_+(x)\psi^{({\textbf{n}})}(x)\,{\rm}dx = 0.
\end{equation}
Orthogonality also holds between any $\psi^{(\textbf{n})}$ and any
ordinary solution $\psi$ with negative energies $E = -\hbar^2
\kappa^2 /(2 m)$, where $\kappa$ is a solution to equation
(\ref{disp-3}).

Finally, the ordinary solutions (\ref{sol-1}), where
$k$ satisfies the dispersion relation (\ref{disp-21}), are
orthogonal to each other for different values of $k$.  For
instance, including subscripts to indicate the $k$-dependence of
wave functions, $\psi_{k,\pm}(x)$ and normalization constants
$C_k$, we find by direct calculation, that when $k \neq k'$,
\begin{eqnarray}\label{}
  \int_0^L \psi_k(x) \psi_{k'}(x)\,{\rm d}x &=&
  \int_0^\ell \psi_{k,-}(x) \psi_{k',-}(x)\,{\rm d}x +
  \int_\ell^L\psi_{k,+}(x) \psi_{k',+}(x)\, {\rm d}x \nonumber\\
    & = & \frac{C_k C_{k'}}{2}(A - B), \nonumber
\end{eqnarray}
where, doing the integrals and rearranging,
\begin{eqnarray}\label{}
    A &=& \frac{1}{(k-k')}\big[\sin (k(L-\ell))\sin
    (k'(L-\ell))\sin((k-k')\ell)\nonumber\\
      &  + & \sin(k \ell) \sin(k' \ell) \sin ((k-k')(L-\ell))\big]\nonumber\\
     \mbox{and}\quad B &=& \frac{1}{(k+k')}
     \big[\sin(k(L-\ell))\sin(k'(L-\ell))\sin((k+k')\ell) \nonumber\\
      & +& \sin(k \ell) \sin(k'\ell) \sin((k+k')(L-\ell)) \big].\nonumber
\end{eqnarray}
Then, by using the eigenvalue equation (\ref{disp-21}) and
trigonometric identities one finds that
\[ A = \frac{\Lambda}{2}\sin(k L)\sin(k' L) = B, \]
so that, when $k \neq k'$,
\begin{equation}\label{orthnorm}
    \int_0^L \psi_k(x) \psi_{k'}(x)\,{\rm d}x = 0.
\end{equation}
\section{weak coupling}\label{weak}
The need to include both the nodal and ordinary solutions to make up
a complete set can be seen by considering the limit of weak
coupling for which $|\lambda|$ is small and $|\Lambda|/L$ large.
In that limit one would expect, whatever the value of $\ell \in (0,L)$, that solutions
(\ref{sol-1}) and their energies would approach those of a free
particle in a well, Eqs. (\ref{well-1}) and (\ref{energy-1}),
no more and no less.

Consider solutions with positive energies. For arbitrary $\ell \in (0,L)$,
in addition to the nodal states---which exist
only when $\ell/L$ is rational---we have the ordinary states. For
them $kL/\pi$ is not an integer and the dispersion relation can be
written, say, as (\ref{disp-21}). Solutions for $kL$ to either equation
are intersections of a straight line passing through the origin with slope $\Lambda/L$,
large positive for attraction or large negative for repulsion. These intersections
occur only near any divergences of the right-hand side (RHS) of Eq. (\ref{disp-21}), i.e.
near the zeroes of $\sin(kL)$.  Then, setting $kL = N \pi + \epsilon$, where $\epsilon$
is small but of either sign, shows that near these zeroes we have, approximately,
\begin{equation}\label{LB-2}
   \mbox{RHS} \simeq \begin{cases}
   -(2/\epsilon)\sin^2( N \pi \ell/L) + \sum\limits_{n\geq 0} a_n(N)
   \, \epsilon^n\quad(\mbox{for}\,\,N=1,2,3 \ldots)\\[.4cm]
   \,\,\,\,\,(\epsilon/2)(1- \mu^2) + b(\epsilon)\quad(\mbox{for}\,\,N = 0,\,\,\epsilon<
   \pi),
    \end{cases}
\end{equation}
where $b(\epsilon)$ is monotonically increasing as a function of
$\epsilon$ in the interval $[0,\pi)$ and $\mu = 2 (\ell/L)-1$.

Suppose first that $ \ell/L $ is irrational.  In that case the
function $\sin(N\pi \ell/L)$ doesn't vanish for any non-zero
integer $N$. Then, using the first of these two equations in
(\ref{disp-21}), with $kL=N \pi + \epsilon$, gives (for
$N=1,2,3,\ldots$),
\[\epsilon = -2 \frac{\sin^2(N \pi \ell/L)}{N \pi (\Lambda/L)} +
\mbox{terms in higher inverse powers of} \,\, ( \Lambda/L). \]
Using the second of equations (\ref{LB-2}) in (\ref{disp-21})
verifies that there is no solution in this limit for $N=0$.  To
summarize: when $\ell/L$ is irrational, solutions for $k L$ to
(\ref{disp-21})---or, what is equivalent, (\ref{LB-1})---will, for
ever weaker coupling ($|\Lambda|/L
\rightarrow \infty$), approach $N \pi$ for attraction or repulsion, where $N= 1,2,3\ldots$, so
making up the full set of energies for a free particle in a well.
Also, it is easy to see, in the limit $kL = N \pi +\epsilon$ with
$\epsilon \rightarrow 0$, that solutions (\ref{sol-1}) approach
those of the free particle in a well, Eq. (\ref{well-1}).

If instead $\ell/L$ should take the rational value $p/n$, which we
assume has been reduced by eliminating all common factors, then
$\sin(N \pi \ell/L)$ will vanish when and only when $N = j\,n$,
$j=1,2,3\ldots$. There are in the limit, therefore, no ordinary
solutions for these values of $N$, but we do have the nodal
states, Eqs. (\ref{sol-3}) and (\ref{sol-31}). In addition to
these nodal states, as we know from the argument just given for
irrational $\ell/L$, for large $|\Lambda|/L$ there exist ordinary
solutions for all other integral values of $N$ again making up the
full set of solutions for a free particle in a one-dimensional well.

For negative energies, similar arguments show that when $|\Lambda|/L$ is large
there are no solutions to either of Eqs. (\ref{disp-3}).
\section{strong coupling}\label{strong}
Now consider, for any $\ell \in (0,L)$, the strong coupling limit,
for which the magnitude of $\Lambda/L$ is small.
Graphical argument shows that in this case, and only for attraction ($\lambda > 0$),
there is a single state of negative energy. This is easily verified by considering
Eq. (\ref{disp-3}) for small positive $\Lambda/L$. For
any value of $\mu$ between $-1$ and $+1$ the right-hand side of
that equation is positive and monotonically increasing as a
function of positive $\kappa L $, from zero asymptotically to the
value +1. Thus, the single solution approaches the
value $\kappa \Lambda = 1$, giving the limiting energy value $E = - \hbar^2
\kappa^2/(2 m) = - E_B$.

Except for the single bound state when the potential is attractive, for
strong coupling the behavior of the oscillatory positive energy solutions to
Eq. (\ref{schrod}) is controlled to some extent by the product $\lambda\,\psi(\ell)$
in the potential energy term, for which one expects
that whatever its sign, as the magnitude of $\lambda$ increases
any ordinary solution evaluated at the potential, $\psi(\ell)$,
will decrease in compensation. Of course when $\ell/L$ is
rational, the corresponding nodal solutions strictly vanish at the
delta function and so are independent of the magnitude of $\lambda$, so
long as it is non-zero.

For the positive energies, when  $|\Lambda|/L$ is small, we can say, from equation (\ref{dispeq}),
that $\sin(k \ell) \sin(k(L-\ell))$ is also small, so that, whatever the sign of $\lambda$,
\begin{equation}\label{bigLambda}
   \mbox{either (a)}\quad k \ell= n_1 \pi+ \varepsilon_1
    \quad\quad \mbox{or (b)}\quad k (L-\ell)=n_2 \pi+ \varepsilon_2,
\end{equation}
where $n_1$ and $n_2$ are any positive integers and
$\varepsilon_1$ and $\varepsilon_2$ are small and vanish as
$|\Lambda|/L \rightarrow 0$. Note that, in
the limits of vanishing $\varepsilon_1$ and $\varepsilon_2$, cases
(a) and (b) cannot both hold, for then, according to
(\ref{sol-1}), the wave function would strictly vanish.

Suppose first that $\ell/L$ is irrational. Then the two cases given
in (\ref{bigLambda}) generate, in the limit, distinct energies and
do not overlap for any values of $n_1$ and $n_2$: If this were not
so, then for some integers $n_1$ and $n_2$ we would have $kL = n_1
\pi$ and $k(L- \ell) = n_2 \pi$ for the same value of $k$ so that
$\ell/L$ would be rational.

Cases (a) and (b) of (\ref{bigLambda}) give, in the limit of
vanishing $\Lambda/L$, nothing other than the energy levels of a
free particle in separate wells of width $\ell$ and $(L-\ell)$. We
can underline this by deriving the wave functions in the strong
coupling limit. For case (a), using in Eq. (\ref{sol-1}) the
first of expressions (\ref{bigLambda}) and taking the limit
$\varepsilon_1 \rightarrow 0$, gives
\begin{equation*}\label{duh}
    \psi(x)\rightarrow
    \begin{cases}
    0\quad\mbox{for}\quad \ell \leq x \leq L \\
    \sqrt{2/\ell}\, \sin(n_1 \pi x/\ell)\quad\mbox{for}\quad
    0 \leq x \leq \ell.
    \end{cases}
\end{equation*}
And, similarly, for case (b): in the limit $\varepsilon_2
\rightarrow 0$ these states approach the ordinary well states constrained
to the interval $[\ell,L]$ with quantum number $n_2$. That the
energies approach those of free particle motion in separate wells
of widths $\ell$ and $L-\ell$ as $|\Lambda|/L\rightarrow 0$ (when
$\ell/L$ is an irrational number) has been pointed out by
Joglekar \cite{jog}.

There are subtleties when $\ell/L$ is rational, with $\ell/L =
p/n$, where we assume as usual that all common factors have been
eliminated from numerator and denominator. In that case, for all
values of $\Lambda$, we have the nodal solutions, Eqs.
(\ref{sol-3}) and (\ref{sol-31}), with wave numbers given by $kL =
j n\pi$, where $j = 1,2,3\cdots.$ And, as before, in addition to
those we have the ordinary solutions for which $\sin(k \ell)
\sin(k(L-\ell))$ is small. For these, from Eq. (\ref{bigLambda}),
\begin{equation}\label{ratB}
    \mbox{either (a)} \quad kL=n_1 \pi\, (n/p) + \varepsilon_a
    \quad\quad \mbox{or (b)}\quad kL=n_2 \pi\, n/(n-p) + \varepsilon_b,
\end{equation}
where $\varepsilon_a$ and $\varepsilon_b$ are small. In this case
possibilities (a) and (b) are both satisfied whenever $n_2/n_1 =
(n-p)/p\,.$ But that instance gives---as we have said---a
vanishing wave function and so must be ruled out as a solution.
Another property is that when $n_1 = j p$ or $n_2 = (n-p)j$, where
$j=1,2,3 \cdots$, they are converging to nodal state energies. But
note that, however small $|\Lambda|/L$ may be, so long as it is
non-zero these ordinary states are distinct from nodal states and
orthogonal to them. In summary: when $\ell/L=p/n$ (reduced) then
the set of limiting values for $kL$ as $|\Lambda|/L$ approaches zero
comprises the nodal values together with all distinct
(non-duplicating) limits in Eq. (\ref{ratB}).

\begin{figure}[h]
\begin{center}
\includegraphics{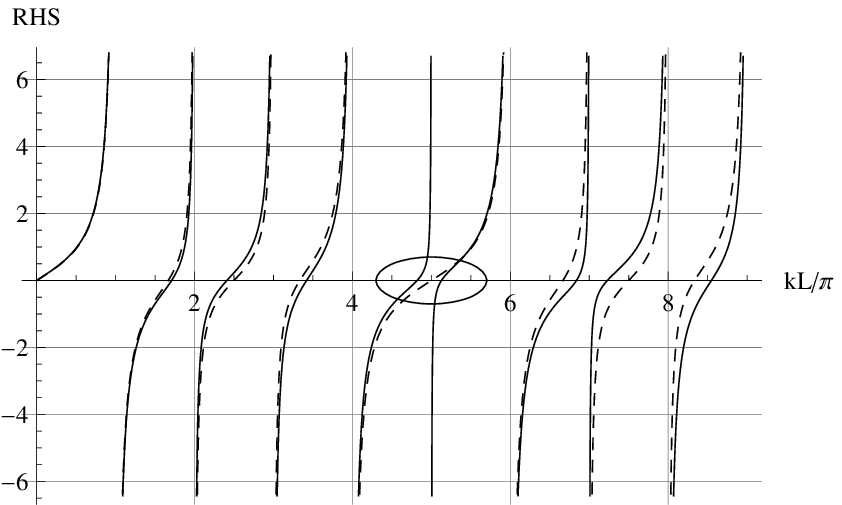}
\caption{\label{lowLambda}This shows the right-hand side (RHS) of Eq. (\ref{disp-21})
as a function of $kL$ in units of $\pi$ for $\ell/L= 2/5$ (dashed)
and $\ell/L =0.415$ (solid).  Solutions for the ordinary states are intersections
of the curves with a straight line from the origin having slope $\pi \Lambda/L$.
When $\ell/L=2/5$ the lowest nodal
state has a wave number given by $kL/\pi = 5$, which is shared with an ordinary
state (dashed line) in the limit $|\Lambda|\rightarrow 0$. That the dashed curve moves from the value
$5$ shows that this limiting degeneracy is lifted as $|\Lambda|$ moves from
zero. On the other hand, note that even when $\Lambda=0$ this
double degeneracy is lifted by moving the potential away from the
value $2/5$.  The fact that $0.415=83/200$ is also rational does
not effect this basic splitting: Its lowest lying nodal state
occurs well to the right, at $kL/\pi =200$.}
\end{center}
\end{figure}

As an example we have considered the case $\ell/L = 2/5$.  Its
nodal states have wave vectors given by $kL/\pi= 5j$,
$j=1,2,3,\ldots$.  In addition to these we have the ordinary
states.  In the limit $|\Lambda|/L \rightarrow 0$ these are given by
the zeroes of the right-hand side (RHS) of Eq.
(\ref{disp-21}) (and (\ref{LB-1})) with $\ell/L = 2/5$.  This is
shown as the dashed lines in Fig.~\ref{lowLambda} as a function
of $kL/\pi$ for values up to $kL/\pi =9$.  The solid curves show
the same function plotted for the nearby value $\ell/L = 0.415 =
83/200$. Solutions for $kL/\pi$ in equation (\ref{disp-21}) occur
at intersections of the curves with a straight line through the
origin with slope $\Lambda \pi/L$.  Thus the figure shows that in the limit $|\Lambda|/L \rightarrow
0$ the degeneracy between the ordinary state and the lowest nodal
energy at $kL/\pi = 5$ is {\em lifted} by moving the potential. It
also shows that, with $\ell/L = 2/5$ for example, the double
degeneracy at $kL/\pi = 5$ (that occurs {\em in the limit} $|\Lambda|/L \rightarrow 0$)
is split when $\Lambda/L$ is nonzero and of either sign. The zeroes of
the functions shown in Fig.~\ref{lowLambda} can easily be
computed from Eq. (\ref{ratB}).

To explore the rational case $\ell/L = 2/5$ a bit further we have
calculated the lowest energy eigenfunction that is doubly
degenerate in the limit $|\Lambda|/L \rightarrow 0$. The lowest
energy nodal state is, from (\ref{sol-3}), given by
$\psi_{5,1}^{(\textbf{n})}(x) =
\sqrt{2/L} \sin(5 \pi x/L)$. To get the ordinary solution at this
energy we let $kL=5 \pi +\varepsilon$ in Eq. (\ref{sol-1})
and take the limit to find
\begin{equation*}\label{2/5}
    \psi(x)=
    \begin{cases}
    \quad D \sin (5 \pi x/L)
    \quad \mbox{for}\quad (2/5)L \leq x \leq L\\
    - (3/2) D \sin(5 \pi x/L)
    \quad \mbox{for}\quad 0\leq x \leq (2/5)L,
    \end{cases}
\end{equation*}
where $D= 2/(\sqrt{3L})$ is a normalization constant.  In the
limit this ordinary solution vanishes at the delta function
$(x=2L/5)$, suffers a discontinuity of slope there, and is
orthogonal to the nodal solution $\psi_{5,1}^{(\textbf{n})}$.
\section{the special case $\ell = 1/2$}\label{1/2}
The simplest choice is to locate the delta potential precisely in
the middle so that $\ell/L$ is the rational number $1/2$ and $\mu
= 0$. For this symmetric case it turns out that the ordinary and
nodal states are precisely equal in number in the sense that their
energies interleave. That is not the case for all other choices of
$\ell/L$, where the symmetry is lost, but it is still true that
both types are countably infinite in number and that their
inclusion is necessary to make up a complete set.

The nodal wave functions and their energies,
$\psi_{2,\,j}^{(\textbf{n})}(x)$ and $E_{2,j}$, are given by
(\ref{sol-3}) with wave vectors $kL =2 j\,
\pi$, $j = 1,2,3\ldots$, Eq. (\ref{sol-31}). These energies
$E_{2,j}$ are, of course, all positive.

For the ordinary solutions any negative energy eigenvalue would
have $k={\rm i}\kappa$ and $E = -\hbar^2\kappa^2/(2m)$ where, putting
$\mu = 0$ in (\ref{disp-3}), $\kappa$ would satisfy the equation
$(\kappa L/2)(2 \Lambda/L)= \tanh(\kappa L/2)$. A sketch of this
equation with respect to the variable $x=\kappa L/2$ shows that there
are no solutions for $\kappa$ when the interaction is a repulsion, none
for relatively weak attraction ($1 < 2 \Lambda/L $), and but a
single solution for relatively strong attraction ($0 < 2 \Lambda
/L< 1$). In particular in the limit of very strong binding, for
which $2 \Lambda/L$ approaches zero from above, $\kappa \rightarrow 1/
\Lambda$, and the energy approaches $ -E_B$.

For the ordinary solutions at positive energies, acceptable values
for $k L$ must, from (\ref{LB-1}), obey the equation $(kL/2)(2
\Lambda/L)=\tan(k L/2)$. This will
include all the ordinary solutions except when the potential is
relatively strongly attractive, for which $0 < 2 \Lambda /L< 1$
and the lowest energy moves to negative values.

When the interaction is repulsive or relatively weakly attractive
all ordinary states have positive energies, and a sketch reveals that,
for this symmetric case, these solutions for $kL$ interleave between those for the nodal solutions.
In this sense, for the special case $\ell=1/2$ the nodal and ordinary states
are equal in number.

For a strong attraction ($0 < 2 \Lambda /L< 1$) there is but one negative energy state
(\ref{sol-11}). If the potential is repulsive, whatever its strength, then all states have positive
energies. In addition to the
nodal states $\psi_{2,\,j}^{(\textbf{n})}(x)$ we have the ordinary
states.  Of these, should there be attraction, there will be, as
$\Lambda/L \rightarrow 0^+$, the single negative energy state with energy
$-E_B$ and wave function approaching $\Psi_{L/2}$, where (see Eq. (\ref{bc-11}))
\[ \Psi_{L/2}(x) = \frac{1}{\sqrt{\Lambda}}\,
{\rm exp}\left({-\frac{|x-L/2|}{\Lambda}}\right)
\quad \mbox{for}\quad 0 < x < L. \]

For strong repulsion (and for strong attraction apart from the
ground state $\Psi_{L/2}$) all states have positive energies $E=
\hbar^2 k^2/(2m)$. For these states $|\Lambda|/L$
is small and solutions for $kL$ are in the neighborhood of the
zeroes of $\tan\left(kL/2 \right)$, namely $kL = 2 j \pi +
\varepsilon$, where $\varepsilon$ approaches zero as
$|\Lambda|/L \rightarrow 0$ and $j=1,2,3\ldots$. Thus, as
$|\Lambda|/L \rightarrow 0 $, the energies of these ordinary states
converge towards a double degeneracy with each and every one of
the the nodal state values $E_{2,j}$, \,\,Eq. (\ref{sol-3}),
except, for attraction, the lowest state.

It is easy to find the positive energy wave functions for strong
interaction. From (\ref{sol-3}) the nodal states are
$\psi_{2,\,j}^{(\textbf{n})}(x) =
\sqrt{2/L}\, \sin\left(2 j \pi x/L\right)$. Using
$kL = 2 j \pi + \varepsilon$ in Eq. (\ref{sol-1}), taking the
limit $\varepsilon \rightarrow 0$ and normalizing, gives the
following expression for the ordinary states in the limit of
strong interaction:
\begin{equation*}
     \psi_j(x) = \theta(x-(L/2))\, \psi_{2,\,j}^{(\textbf{n})}(x)
     \quad (j = 1,2,3\ldots),
\end{equation*}
where $\theta$ is the step function
\begin{equation}\label{theta}
    \theta(u) =
\begin{cases}
 +1\quad u>0 \\
 -1 \quad u<0\,.
\end{cases}
\end{equation}
These ordinary states have a discontinuous slope at the delta function, are even valued with
respect to reflection about $x = L/2$ and are manifestly
orthogonal to the nodal states which are odd valued about the
middle. In this limit the energy of each ordinary state $\psi_j(x)$ approaches that of the
corresponding nodal state $\psi_{2,\,j}^{(\textbf{n})}(x)$: in
this case of high symmetry all states approach doubly degeneracy in the
limit of strong interaction. However we note that
for all finite strengths of the contact potential all energies of this
one-dimensional system are, of course, non-degenerate.
\section{ground state}\label{gs}
We considered in subsection \ref{1/2} solutions to Eqs.
(\ref{LB-1}) and (\ref{disp-3}) with $\mu = 0$, i.e. the case
$\ell = L/2$. In that instance the ground state energy is
positive for all strengths of repulsion, $\Lambda < 0$, and also
for relatively weak attraction, $2 \Lambda/L> 1$.  But for
relatively strong binding, for which the inequality $1 > 2\Lambda
/L > 0$ holds, the ground state energy is negative. And, in
particular, in the limit of strong binding, when $\Lambda/L
\rightarrow 0^+$, it approaches $-\hbar^2/(2m
\Lambda^2)= -E_B$, the bound state energy of the free system,
Eq. (\ref{ebee}).

When the force center is not in the middle, then $\mu$ does not
vanish and a re-analysis is required. For that case, whatever the
details may be, we can say from the start that for either sign and
any magnitude of $\Lambda$, because Eqs. (\ref{LB-1}) and
(\ref{disp-3}) are symmetric with respect to $\mu$ about $\mu=0$,
as a function of $\ell$, any ordinary state energy eigenvalue must
be symmetric about $\ell = L/2$. It is easy to see from
(\ref{sol-3}) that this symmetry also applies to the nodal states.

Consider, in particular, configurations with the delta function
located near to the left wall, say, with $\ell =
\epsilon$, where $\epsilon$ is small and positive. As $\epsilon$ decreases, the
nodal solution energy eigenvalues--- see Eq. (\ref{sol-3})---will
rapidly increase to large positive values. To
see this, note that with $\ell/L = p/n$, $n$ must increase without
bound as $\epsilon \rightarrow 0^+$, taking with it the
accompanying energies $E_{n,j} \sim (j\, n)^2$, $j=1,2,3,\cdots$.

As for the ordinary solutions, with $\ell = \epsilon$ and
$\epsilon\rightarrow 0^+$, we can say that as the delta function
approaches a wall there are no negative energies. To see this,
expand Eq. (\ref{disp-3}) for small $\epsilon$ to get
$\tanh(\kappa L)/(\kappa L) \simeq -2
\epsilon^2/(\Lambda L)$.  In the limit $\epsilon
\rightarrow 0^+ $ this has no solutions for any finite value of
$\kappa$. So all energy eigenvalues are non-negative when the
delta function potential is near to either wall.

For arbitrary $\mu \in (-1, 1)$, a sketch allows us to predict the existence
of ordinary solutions for small positive and negative energies,
the transition between which occurs at $k=0$. The second of
Eqs. (\ref{LB-2}) (which applies when $kL$ is small) and the
graphical picture suggests the following: that for small $kL$ and
an attractive potential ($\Lambda > 0$) there is a solution for
small real valued $kL$ provided the slope $\Lambda/L$ is greater
than $(1-\mu^2)/2$; that $k$ vanishes when $\Lambda / L = (1-
\mu^2)/2$; that there is a negative energy solution with $k = {\rm i}\kappa$ when
$\Lambda/L$ is less than $(1- \mu^2)/2$.  From Eq.
(\ref{LB-1}) $\mu = 2\ell/L-1$, so we can say that as a function
of $\ell/L$, the ground state energy passes through zero when
$\Lambda/L = [1- \mu^2]/2 =2 (1 -\ell/L) \ell/L$, or
\begin{equation}\label{kvan}
    \ell/L =(1/2) \left(1 \pm \sqrt{1 - 2 \Lambda/L}\right).
\end{equation}
Note that, because $\ell/L$ is restricted to the interval $(0,1)$,
there can only be zeroes of energy provided $0 < 2\Lambda/L < 1$,
i.e. for relatively strong attraction.

Finally, consider the ordinary solutions for values of $\ell$
near, say, the left wall so that $\ell = \epsilon$, where
$\epsilon$ is small. Expanding Eq. (\ref{disp-21}) to lowest
order and collecting terms gives $\tan(kL)/(kL) \simeq -2
\epsilon^2/\Lambda L$, so that in the limit $\epsilon \rightarrow
0$, $kL \rightarrow N \pi$ $(N = 1,2,\cdots)$, or $E \rightarrow
\hbar^2 N^2 \pi^2/(2m L^2)$. Thus, in this limit the energies
approach those of a free particle in a well, Eq.
(\ref{energy-1}). To see how this limit is approached as $\ell
\rightarrow 0^+$ we can set $kL = N
\pi + \Delta$, where $\Delta$ is small.  Then $\tan(kL)/kL = \tan
\Delta/(N \pi + \Delta) \simeq \Delta/(N
\pi)$, so that $\Delta/(N \pi) \simeq -2 \epsilon^2/(\Lambda
L)$ or $\Delta \simeq -2 N \pi \epsilon^2/ (\Lambda L)$. The
corresponding energy is
\begin{equation}\label{wall}
    E = \frac{\hbar^2}{2mL^2}(N \pi + \Delta)^2 \simeq
\frac{\hbar^2}{2m}\left(\frac{N \pi}{L}\right)^2
\left(1 - \frac{4 \epsilon^2}{L \Lambda}\right),\quad
N=1,2,3\cdots.
\end{equation}
This result holds for either sign of $\Lambda =
\hbar^2/(m \lambda)$. The conclusion, then, is that when the delta
function is near either wall, as a function of distance away, the
energies are parabolic in form with zero slope at the wall, and
they are a local maximum (minimum) for an attractive (repulsive)
interaction.

\begin{figure}[h]
\begin{center}
\includegraphics{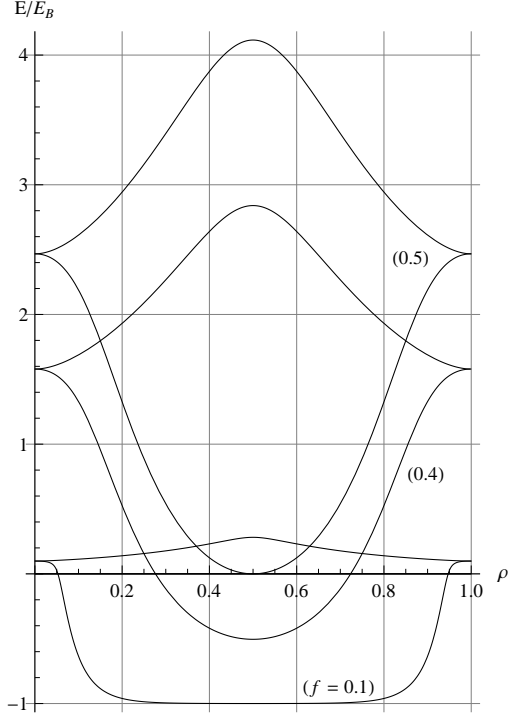}
\caption{\label{single}The result of numerical calculations
(for values $(0.1,0.4,0.5)$ of the parameter $f=\Lambda/L$) for
the dependence of the `Relative Energy' $E/E_B$ of the lowest
energy eigenvalue versus $\rho =\ell/L$ locating the delta
function within a box of side $L$.  The curves with minima are for
an attractive potential $(\lambda > 0)$.  Their partner curves,
with maxima, correspond to a repulsive interaction of the same
strength.  The energy of the single bound state of an attractive
potential of strength $ |\lambda |$ is $E = - E_B$, Eq.
(\ref{ebee}), and $\Lambda$ is the spatial decay length for that
state, Eq. (\ref{biglam}).}
\end{center}
\end{figure}

The lowest, ground state, energy corresponds to the choice $N=1$
in Eq. (\ref{wall}). Numerical calculation gives the
dependence of energy eigenvalues on $\ell$ as it is varied from
$0$ to $L$. All energies above the ground state are positive so
for them one must solve either of the pair (\ref{disp-21}) or
(\ref{LB-1}). For the ground state, when $2
\Lambda/L > 1$, these equations pertain, but when $2 \Lambda/L <
1$ Eq. (\ref{disp-3}) holds. Their solutions join smoothly at
zero energy, for which $k$ vanishes. This happens at values of
$\ell$ given by Eq. (\ref{kvan}). Results for the ground
state are shown in Fig.~\ref{single} where the ratio $E/E_B$
versus $R = \ell/L$ is plotted for the three values
$(0.1,0.4,0.5)$ for the parameter $f = \Lambda/L$.  We have done
this for attractive and repulsive delta function potentials having
the same strength $| \Lambda |$. For the former (latter) case
$\Lambda> 0$ ($\Lambda< 0$) and the the energy eigenvalues have
their minimum (maximum) value at $f=1/2$.

Generally, for the case of attraction, should the wall separation
$L$ increase to values large with respect to the characteristic
length $\Lambda$ one expects that the ground state energy would
approach $-E_B$, Eq. (\ref{ebee}).  Indeed in the limit
$\kappa L \rightarrow \infty$ Eq. (\ref{disp-3}) becomes
$\kappa \rightarrow 1$, provided $0 < \ell < L$ so that $\kappa
\rightarrow 1/\Lambda$ and $E \rightarrow - \hbar^2/(2m
\Lambda^2)= - E_B$. Note, however, that however large $L$ may be,
according to Eq. (\ref{wall}) with $n=1$, in the limits $\ell
\rightarrow 0^+$ and $\ell \rightarrow L^-$ the wall dominates so
that the energy approaches $\hbar^2 \pi^2/(2m L^2)$, the lowest
energy of a free particle in the well. This is consistent with
Fig.~\ref{single}. In the other extreme of small $L$, the walls
dominate, whatever the value of $\ell$. In this limit
$|\Lambda|/L$ is large, which is equivalent to weak coupling. For
this limit the analysis around the first of Eqs. (\ref{LB-2})
shows that $kL \rightarrow \pi$ as $|\Lambda|/L \rightarrow
\infty$, so that in the limit $L\rightarrow 0$ we find that $E \rightarrow \hbar^2
\pi^2/(2m L^2) = \pi^2 f^2 E_B$, the lowest energy of a
free particle in a well.
\section{concluding remarks}\label{finish}
The model analyzed here is simple and exactly solvable, but we are
unaware of any complete published analysis. If we think of an
attractive delta function as a very simple model of the nucleus of
a relatively massive atom, then along the lines of the
Born-Oppenheimer approximation we might suppose that its position
$\ell \in (0,L)$ can be varied as a parameter.  Thus the model
suggests, reasonably, that for an attractive interaction the
configuration of lowest energy is for the `nucleus' to avoid the
walls. This agrees with the perturbation calculations of Changa et
al \cite{changa} for hydrogen in a hard spherical cavity. Note
that as the well width is decreased the uncertainty principle
demands that eventually the kinetic energy term of the
Schr\"{o}dinger equation is the dominant contribution to the
energy. As we pointed out in Sec.~\ref{gs} this also applies for a
well of any width should the fixed potential be located
sufficiently close to either wall.

We have shown that the energy levels and wave functions for choice
$\ell = L/2$ follow as the limit of a `top hat' model potential
having height $H$ and width $W < L$ where the `area' $HW$ is held
constant and $W \rightarrow 0^+$, and it is reasonable to conjecture
that the same result applies for off-center top hat potentials and, more generally, to
any sequence of model potentials which approaches the delta
function as a limit.

As a limit of sequences of non-singular potentials, the choice
$v(x) = - \lambda \delta (x- \ell)$ does require care in its
handling. In particular the behavior of the nodal energies, Eqs.
(\ref{sol-3}) and (\ref{sol-31}), is somewhat chaotic as $\ell$ is
varied. For instance, for any given `rational' position $\ell/L =
p/n$, the lowest nodal energy is proportional to $n^2$, so that the
lowest such energies occur, in increasing order, at relative
position $\ell/L=1/2$ and then at $\ell/L = (1/3,1/4,1/5,\cdots)$ to the left of the midpoint or,
symmetrically, at points $\ell/L =(2/3,3/4,4/5,\cdots)$ to its
right. But should $\ell/L$ be moved through nodal points {\em
arbitrarily close} to any one of these points then the lowest
associated energy sweeps through chaotically large, even infinite,
values. Nevertheless, we emphasize that, for rational values of
$\ell/L$ a complete set of solutions requires both the nodal and
ordinary ones.  By contrast, the ordinary solutions have energies
which are solutions to Eqs. (\ref{disp-21}) or (\ref{LB-1}). They
are smooth continuous functions of position $\ell \in (0,L)$. For
instance this is true for the ground state discussed in Sect.
\ref{gs}.

To conclude the paper we should like to make a few remarks of a purely mathematical nature concerning the model, most of which
remains to be done. First we prove that the Hamiltonian, properly defined mathematically, is self adjoint. As the reader will
know, self adjointness is that subtle property of a symmetric (hermitian) operator that guarantees that the operator can be
exponentiated. For the Hamiltonian, this condition is necessary and sufficient to guarantee a unique time translation group. For
our formal Hamiltonian, which we here denote
$$S=-\frac{\hbar^2}{2m}\,\frac{d^2}{dx^2}-\lambda \delta(x-\ell),$$
the process begins by defining a domain of definition for its interpretation as a sesquilinear \textit{form}. We choose as domain
$\mathcal{D}$ the set of functions $f\in L^2(0,L)$ in the Hilbert space which belong to the domain of the kinetic operator and,
in addition, have a first derivative $f^\prime$ which satisfies the discontinuity boundary condition at $x=\ell$. This implies
that $f^\prime\in L^2(0,L)$. We can then consider the form
$$\pair{g}{f} :=\int_0^L \clx{g(t)}\, [Sf](t)\,dt$$
for all $f$ in the domain $\mathcal{D}$. By a simple calculation, involving no more than integration by parts, it follows that
the associated quadratic form is lower bounded,
\begin{equation}
\pair{f}{f} = \frac{\hbar^2}{2m} \M{f^\prime}^2-\lambda \m{f(\ell)}^2  \,\ge
 -\lambda\,\sup_{0\le x\le L}|f(x)|^2\, >-\infty\,.
\end{equation}
It is now standard that the form can be extended to a closed form which defines a self adjoint operator, the Friedrichs extension
of $S$.  It is this extension that we take as the mathematically proper Hamiltonian $H$. For details of the extension we
recommend Davies \cite{EBD1} and \S124 of Riesz and Sz.-Nagy \cite{RN}.

This leaves the following to prove: the spectrum of $H$ consists only of eigenvalues, indeed of those eigenvalues we have
obtained earlier and only those. We firmly believe these to be true, and that, in consequence, for a given $\ell$ and $\lambda$
the eigenfunctions we obtained constitute an orthonormal basis for $L^2(0,L)$. We remark that were we to independently prove that
the eigenfunctions were complete then the eigenvalues we found would comprise the spectrum of $H$.

%

%
\newpage
\section*{Figures}
\end{document}